\documentclass{elsart}

\usepackage{natbib}

\usepackage{epsfig}

\usepackage{amssymb}

\begin{document}

\begin{frontmatter}



\title{Removing noise from correlations in multivariate stock price data.}
\author{Przemys\mbox{\l}aw Repetowicz }
\ead{repetowp@tcd.ie}
\author{Peter Richmond}
\ead{richmond@tcd.ie}
\address{Department of Physics, Trinity College Dublin 2, Ireland}


\author{}

\address{}

\begin{abstract}
This paper examines the applicability of Random Matrix Theory 
to portfolio management in finance.
Starting from a group of normally distributed stochastic processes 
with given correlations we devise an algorithm for removing noise
from the estimator of correlations constructed from measured time series.
We then apply this algorithm to historical time series for the 
Standard and Poor's
500 index.
We discuss to what extent the noise can be removed and
whether the resulting underlying 
correlations are sufficiently accurate 
for portfolio management purposes. 
\end{abstract}


\end{frontmatter}

\section{Introduction}
The application of Random Matrix Theory to the analysis of correlations 
of financial time-series has been widely studied by
\citet{Sengupta,Laloux,Bouchaud,Plerou,PlerouI,PlerouII,
Burda,Mantegna,Utsugi,
Malevergne,PafkaI,PafkaII} in the case of Gaussian matrices and by
\citet{Galluccio,Jurkiewicz,Sornette} in the non-Gaussian case.
The non-Gaussian case is more realistic but also more difficult.
One of the reasons of these difficulties is that it is not 
straightforward to use variances and covariances  as 
a measure of risk in the non-Gaussian case as it is in the Gaussian case.
\citet{Sornette} approaches this problem by
mapping asymptotically power-law distributed
assets in a non-linear way onto Gaussian variables
whose covariances provide a new measure of dependence between non-normal returns.
 
In this paper, however, we are going to forget that price returns 
have ``fat'' tails, 
as it was first suggested by \citet{Mandelbrot,EugFama} 
and then proved by \citet{Gopikrishnan}
in a thorough statistical analysis,
and test the case of Gaussian Random matrices. 
Here it is the correlation matrix that measures the 
mutual dependence of normally distributed financial time series
and this is the quantity we will investigate.

In this ensemble of random matrices there are several universal quantities,
i.e. quantities which are the same for both a particular random matrix 
and the average over the whole ensemble. 
In principle then we can measure these quantities for a financial 
correlation matrix, compare them with those from the ensemble,
which will be defined in the next section, and then extract information 
about the underlying correlations between stocks.

One universal quantity is the eigenvalue distribution.
The density of eigenvalues for financial correlation 
matrices seems to be divided into
two or more clearly separated parts (see  Fig. \ref{fig:Correl100}).
The first part, comprising the majority of the spectrum, corresponds
to small eigenvalues. The other parts contain very few eigenvalues 
that are generally much larger than those from the first group.
The components of these other groups are clearly separated from 
the first group and from one another.
Since eigenvalue spectra in the ensemble  have a similar structure
the method consists in modifying the spectrum, 
replacing each group of eigenvalues by groups
of degenerate eigenvalues and reconstructing
the correlation matrix from that modified spectrum and the eigenvectors
of the original financial correlation matrix. 

This method is valid only under certain assumptions.
\begin{enumerate}
\item The financial time series should comprise Gaussian stochastic processes and
\item the underlying matrix of stock correlations should have 
      only a small number of distinct eigenvalues 
$\lambda_i$ for $i=1,\ldots,\mathcal{P}$
that are clearly separated from one another  
$\lambda_1 \ll \lambda_2 \ldots \ll \lambda_\mathcal{P}$.
\end{enumerate}

In this paper we examine the method for 
Gaussian distributed pseudo-random  variables 
conforming to given correlations.
We test the extent to which one can reproduce the underlying correlations
from Gaussian time series of a given finite length.
Furthermore, we apply the algorithm to historical time series of the Standard and Poor's
500 index and compare the results to a different ``denoising'' algorithm suggested by 
\citet{Bouchaud}.

The paper is organized as follows. 
In section \ref{sec:Anal} we briefly discuss the theoretical approach,
in section \ref{sec:fit} we describe the procedure of fitting the theory
to measured data. 

In the sections \ref{sec:simul} and \ref{sec:results} 
we discuss the simulation and 
the results for financial data. 
In section \ref{sec:OptPortf} we discuss whether the amount of noise removed
from the measured correlation matrix of Gaussian processes is sufficient for the 
cleaned  matrix to be used for portfolio management purposes.

We conclude with a brief discussion.

\section{Theoretical formalism\label{sec:Anal}}
Consider a ensemble of normally distributed stochastic processes 
${\bf X} = \left\{ X_{i,t} \right\}$ for $i=1,\ldots,N$ and $t=1,\ldots,T$.
whose joint probability density reads:
\begin{eqnarray}
P\left( {\bf X} \right) D{\bf X} &= &
\prod_{t=1}^T \left( P\left( X_{1,t},\ldots,X_{N,t} \right) \prod_{n=1}^N d X_{n,t} \right) \nonumber \\
&=& {\mathcal N}\exp( -\frac{1}{2} 
\sum_{t=1}^T \sum_{i=1}^N \sum_{j=1}^N X_{i,t} C^{-1}_{i,j} X_{j,t} ) 
\prod_{n,t=1}^{N*T} d X_{n,t} \nonumber \\ 
&=& {\mathcal N}\exp( -\frac{1}{2} Tr \left[ {\bf X^T C^{-1} X } \right] ) D {\bf X}
\end{eqnarray}
with a normalization factor ${\mathcal N}$ that ensures \\
$\int P\left( {\bf X} \right)D{\bf X} = 1$. 

The estimator of correlations in these processes is defined in the usual way:
\begin{equation}
c_{i,j} = \frac{1}{T} {\bf X} {\bf X^T} = \frac{1}{T} \sum_{t=1}^T X_{i,t} X_{j,t} 
\quad i,j=1,\ldots,N
\label{eq:estimator}
\end{equation}

\begin{figure}
\vbox{      \psfig{figure=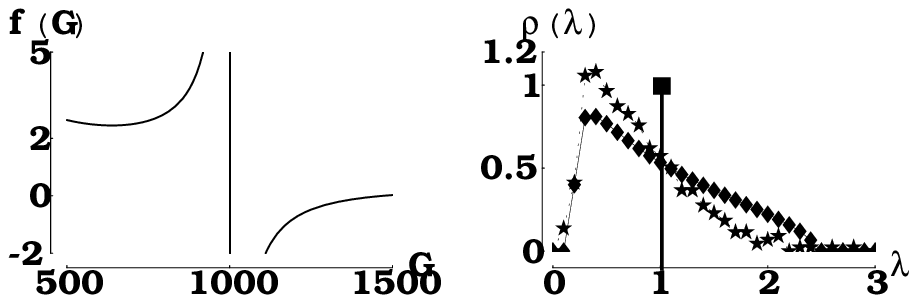,width=\textwidth,angle=0}
            \psfig{figure=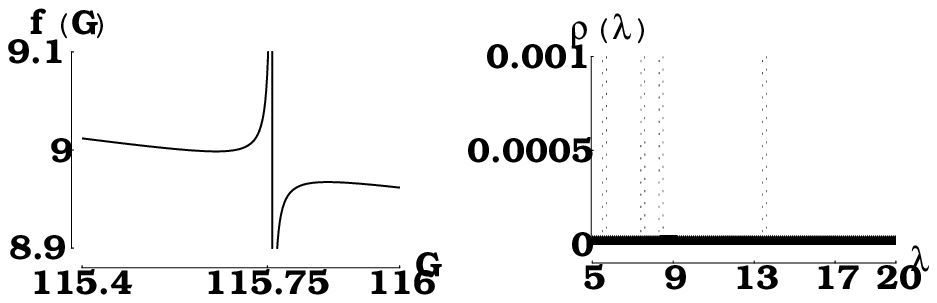,width=\textwidth,angle=0}
            \psfig{figure=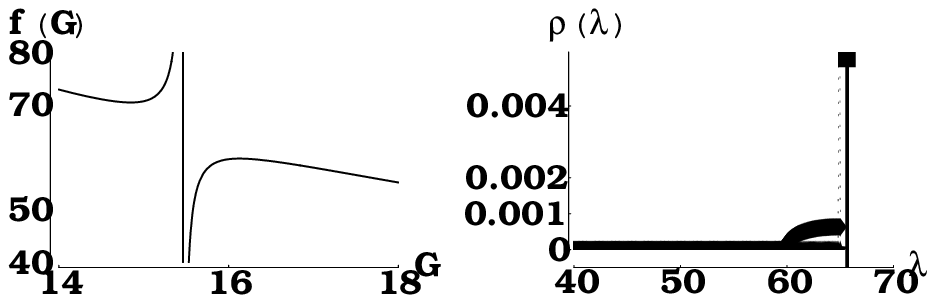,width=\textwidth,angle=0}}
\caption{Illustration of how the imaginary part of the resolvent $\rho_\Lambda(\lambda)$ 
         defined in (\ref{eq:DOEWishart}) and (\ref{eq:ResolventDef})
         splits into several bands 
         when $\mathcal{P} > 1$.
         In each row the Figure on the left-hand side shows a 
         plot of function $z = f(G)$ zoomed out in the vicinity of a singularity.
         On the right-hand side we plot the imaginary part of the resolvent
         $\rho_\Lambda(\lambda)$ 
         as a function of $\lambda$ (marked with diamonds) 
         overlaid with the empirical density of eigenvalues 
         (asterisk). Vertical beams ended with a squares denote eigenvalues $1/\kappa_\xi$ 
         of the cleaned correlation matrix $C_{i,j}$
         and their degeneracies $p_\xi$  
         (compare equation (\ref{eq:ResolventDef})).
         The singularities of the function $z = f(G)$ have been chosen
         in order to achieve a best conformance of $\rho_\Lambda(\lambda)$ 
         to market data of the AA--ZION stocks from the Standard and Poor's group.
         \label{fig:DOEBands}}
\end{figure}

It follows from this definition that the processes $X_{i,t}$ and $X_{j,t'}$ 
with different indices $i\ne j$
are correlated only if the times are the same $t = t'$, i.e. 
\begin{equation}
\left< X_{i,t} X_{j,t'} \right> = C_{i,j} \delta_{t,t'}
\end{equation}
Now we can calculate analytically certain averages over the ensemble
of stochastic processes $X_{i,t}$ called Wishart matrices \citet{Wishart}.
In particular it is possible \citet{Feinberg,Sengupta},
to obtain a closed form expression for the 
trace of a statistically averaged matrix that involves 
an estimator $c_{i,j}$ of the cross correlations.
This resolvent is defined as follows:
\begin{equation}
G(z) = \left< Tr\left[ z - c \right]^{-1} \right>
     = \int Tr\left[ z - \frac{1}{T} {\bf X} {\bf X^T} \right]^{-1} 
       P\left( {\bf X} \right) D{\bf X}   
\label{eq:Greensfct}
\end{equation}
where the integral on the right-hand-side stands for $N\times T$ integrals
over each of the variables $X_{i,t}$ in the limits from $-\infty$ to $\infty$.
The result reads:
\begin{equation}
G(z) = \frac{T}{z - \Sigma(x)} 
\quad\mbox{where}\quad 
\Sigma(z) = Tr\left[ \frac{ {\bf C} }{1 - {\bf C }G(z) } \right]
\label{eq:resolvent}
\end{equation}
In the case when the underlying correlation matrix ${\bf C}$ 
has only few distinct eigenvalues $\sigma_j$ with multiplicities $p_j$ 
for $j=1,\ldots,\mathcal{P}$ the resolvent satisfies an algebraic $(\mathcal{P}+1)$-order 
algebraic equation of the form:
\begin{equation}
z = \frac{T}{G(z)} - \sum_{\xi=1}^\mathcal{P} \frac{p_\xi}{G(z) - \kappa_\xi T}
\quad\mbox{where}\quad \kappa_1 < \kappa_2 < \ldots < \kappa_\mathcal{P}
\label{eq:ResolventDef}
\end{equation}
where $\kappa_\xi = 1/\sigma_\xi^2$ 
The following two constraints are imposed
on the parameters $p_i$: 
$ \sum_{\xi=1}^\mathcal{P} p_\xi = N  \label{eq:Constr1} $
and 
$ \sum_{\xi=1}^\mathcal{P} p_\xi/\kappa_\xi = N $.
These follow from the fact
that the correlation matrix ${\bf C}$ has a dimension $N$ and a trace $N$.
The resolvent is a useful tool for calculating the density of eigenvalues 
\begin{equation}
\rho_\Lambda(\lambda) = \left< \sum_i \delta(\lambda - \lambda_i) \right>
\label{eq:DensityOfEigenvalues}
\end{equation} 
of the estimator of the correlation matrix $c_{i,j}$ in the limit 
$N \longrightarrow \infty$. The density of eigenvalues 
$\rho_\Lambda(\lambda)$
is related to the 
resolvent as follows: 
\begin{eqnarray}
\rho_\Lambda(\lambda) 
&=& \frac{1}{2 \pi} { \mbox{lim}_{\epsilon \longrightarrow 0} }
   Im\left[
    G(\lambda + i \epsilon)  -
    G(\lambda - i \epsilon) 
   \right]
\label{eq:DOEWishart}
\end{eqnarray}  
and is universal, which means
that it doesn't matter whether we perform the average over the 
ensemble of Wishart matrices in (\ref{eq:DensityOfEigenvalues})
or whether we drop the brackets $\left<\right>$ and compute the sum 
over eigenvalues of a particular representative of the ensemble.
Numerically, the density of eigenvalues $\rho_\Lambda(\lambda)$
is computed by solving the algebraic equation (\ref{eq:ResolventDef})
for $G(z)$, taking the imaginary part of the solution and dividing 
it by $\pi$. It is not difficult to realize that 
the imaginary part of the resolvent $\rho_\Lambda(\lambda)$
is different from zero only if $\lambda$ belongs to certain intervals
$\lambda^{(\xi)} - \delta^{(\xi)} \le \lambda \le  \lambda^{(\xi)} + \delta^{(\xi)}$
such that $G(\lambda^{(\xi)}) \simeq  \kappa_\xi T$ (see Fig. \ref{fig:DOEBands}).

For more details of this approach we refer the interested 
reader to original works by \citet{Feinberg,Sengupta}.
We now discuss how to fit the
density of eigenvalues $\rho_\Lambda(\lambda)$ in the ensemble of random matrices
to $\rho_\Lambda^E(\lambda)$  corresponding 
to cross-correlations of financial time series.

\section{The fit algorithm\label{sec:fit}}
As mentioned in the introduction, eigenvalue spectra of cross-correlation matrices 
of financial time series consist of clearly separated groups of eigenvalues 
to be referred to as bands.
Each band corresponds  to one term in the sum on the right-hand-side 
of equation (\ref{eq:ResolventDef}). The parameters $p_\xi$ and $\kappa_\xi$
of the resolvent
are determined, roughly, by the number of eigenvalues in the $\xi$th band and by 
the location of the midpoint of the band.
We calibrate the parameters of the resolvent against market data 
in the following manner.
After computing the eigenvalues of the estimator of correlations 
(\ref{eq:estimator}) we calculate the histogram of eigenvalues and the smoothed density of eigenvalues
$\rho_\Lambda^E(\lambda)$ which is defined as a sum of narrow Gaussians pin-pointed at 
all  eigenvalues, i.e
\begin{equation}
\rho_\Lambda^E(\lambda) = \frac{1}{N} \sum_{i=1}^N \delta_\epsilon\left( \lambda - \lambda_i\right)
\quad\mbox{where}\quad \delta_\epsilon(x) = \frac{1}{\sqrt{2 \pi \epsilon^2}} \exp\{ \frac{x^2}{2\epsilon} \} 
\label{eq:DensEigSmooth}
\end{equation}
The smoothed density $\rho_\Lambda^E(\lambda)$ is used to find the preliminary location of bands,
i.e. their midpoints $1/\kappa_\xi$,
and the numbers of eigenvalues belonging to bands $p_\xi$.
Having determined the parameters we compute the resolvent $G(\lambda)$ 
from equation (\ref{eq:ResolventDef}) 
by finding complex roots of a polynomial equation 
(computations are carried out with multiple precision due to high susceptibility
of polynomial-roots to errors in coefficients)
and then taking the imaginary part to get the $\rho_\Lambda(\lambda)$ 
in the ensemble of Wishart matrices.
Finally we compute the optimal parameters $\kappa_\xi$,
subject to the constraints,
by minimizing the deviation between $\rho_\Lambda^E(\lambda)$ and  $\rho_\Lambda(\lambda)$
using a down-hill simplex method -- a multi-dimensional minimization routine  
from Nelder and Mead \citet{NumericalRecip}.
The parameters $p_\xi$ are not altered in the procedure.
The minimization is carried out in $\mathcal{P}-1$-dimensional space.
Specifically we minimize the sum of relative differences of moments:
$\delta = \sum_{j=1}^8 
\left| \left< \lambda^j \right> - \left< \lambda^j \right>^E \right|/\left< \lambda^j \right>^E$
where $\left< \lambda^j \right> =  \int_0^\infty \lambda^j \rho_\Lambda(\lambda) d\lambda$
and the ``empirical'' mean $\left< \lambda^j \right>^E$ refers to 
the density of eigenvalues of the measured estimator of correlations.

From the optimal parameters $p_\xi$ and $\kappa_\xi$ we reconstruct the spectrum
by replacing the eigenvalues of the estimator $c_{i,j}$ 
by $1/\kappa_1$ with degeneracy $p_1$, $1/\kappa_2$ with degeneracy $p_2$ and so on until
$1/\kappa_\mathcal{P}$ with degeneracy $p_\mathcal{P}$.
Having done that we construct the ``cleaned'' correlation matrix $c^{clean}_{i,j}$ 
by multiplying a diagonal matrix $D$ with the new eigenvalues on the diagonal by 
two matrices on the right and left side respectively.
On the right-hand side we have a matrix $V = \left( \vec{v}_1, \ldots, \vec{v}_N \right)$ 
whose rows are eigenvectors of $c_{i,j}$,
on the left-hand side we have the transposition $V^T$.
It means that $c^{clean}_{i,j} = \sum_{p} \lambda_p V_{p,i} V_{p,j}$.
Finally we check the ``degree of cleaning'', which we define as 
the deviation between the estimator and the cleaned estimator 
$\left| c - c^{clean} \right|$.

\section{Testing the algorithm by simulation\label{sec:simul}}
In order to check the reliability of the output of the algorithm applied to real data
we perform Monte-Carlo simulations.
At each time step $t$ we generate $N$ correlated Gaussian processes $X_i(t)$ with 
given cross-correlations $C_{i,j}$.
This is done by generating at each time step 
$N$ uncorrelated Gaussian processes $Y_i(t)$,
by means of the Box-Mueller algorithm \citet{NumericalRecip},
and building from them $N$ linear combinations 
$X_i(t) = \sum_{p=1}^N V_{i,p} \sqrt{\lambda_p} Y_p(t)$
with coefficients that depend on eigenvalues $\lambda_p$ and eigenvectors $V_{i,p}$
of the underlying correlation matrix $C_{i,j}$.

\begin{figure}
\centerline{\psfig{figure=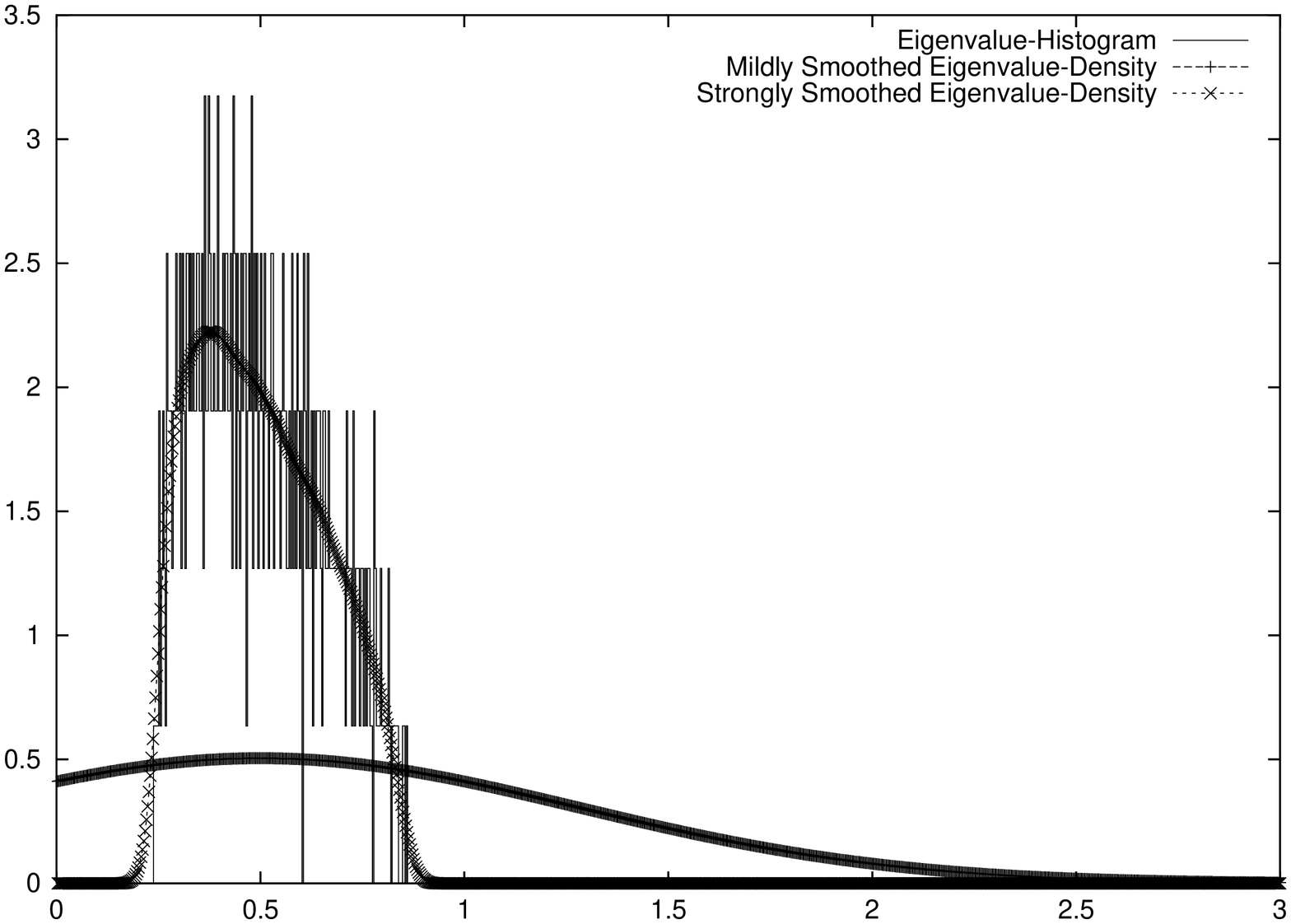,width=0.5\textwidth,angle=-90}}
\centerline{\psfig{figure=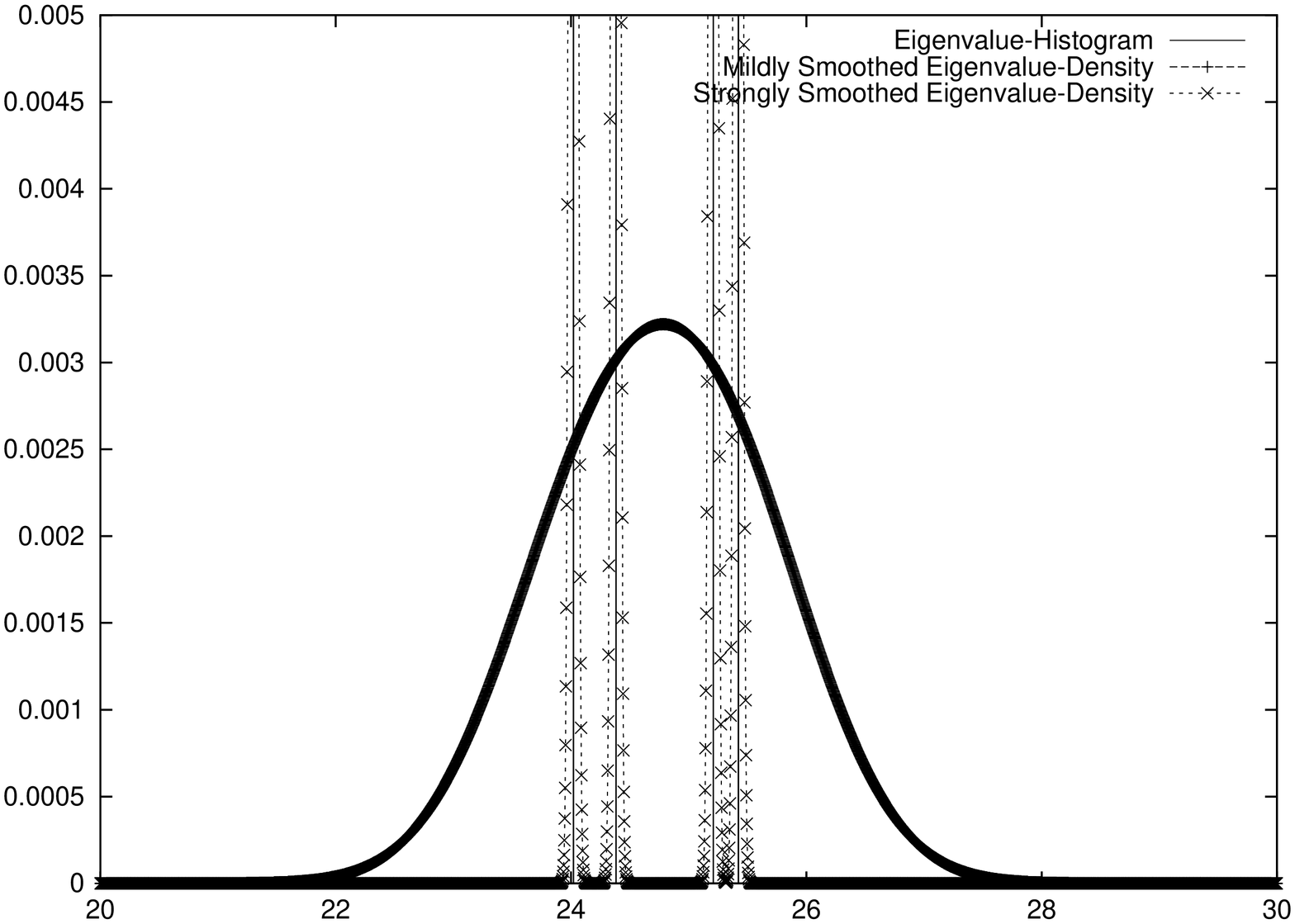,width=0.5\textwidth,angle=-90}}
\centerline{\psfig{figure=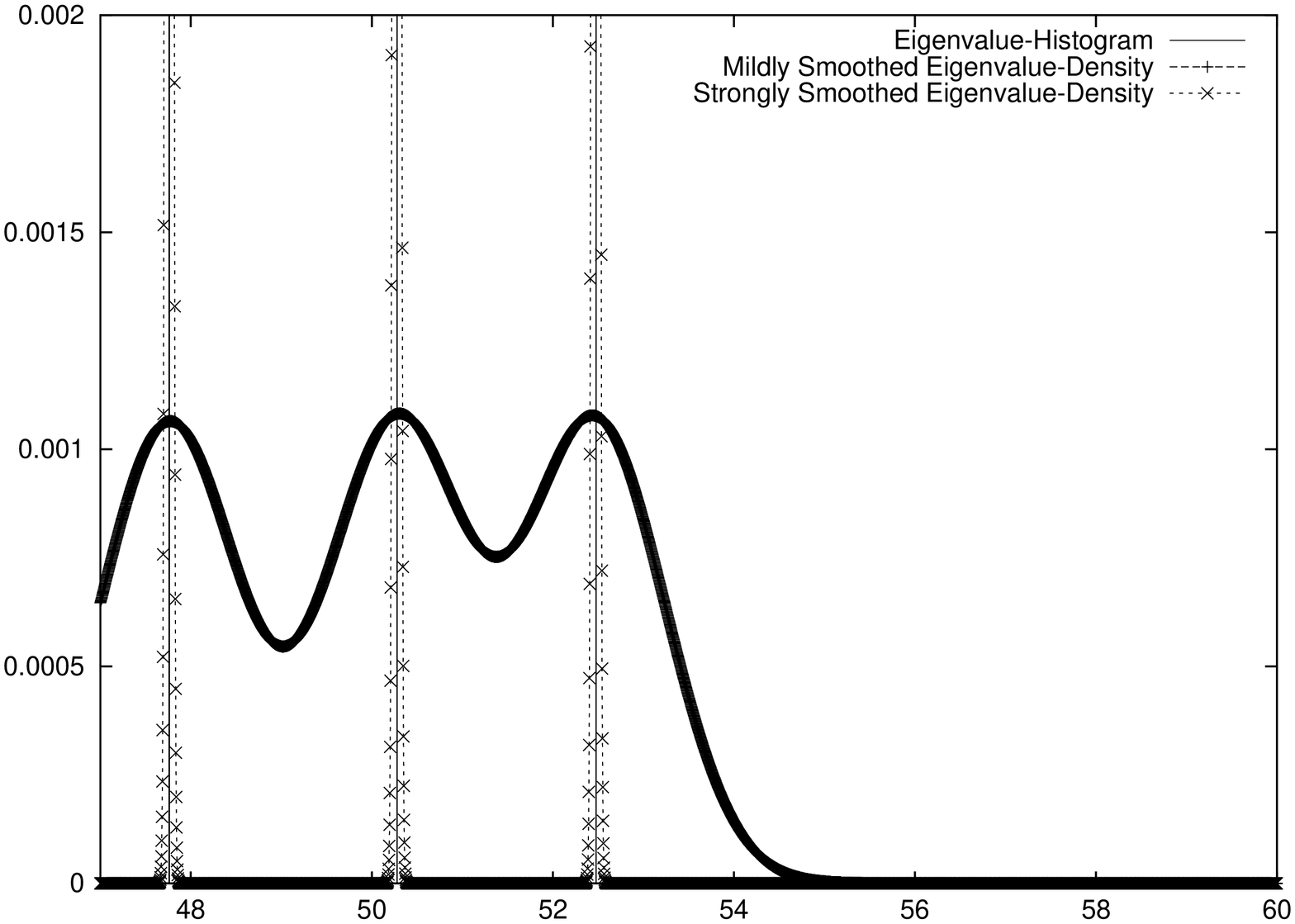,width=0.5\textwidth,angle=-90}}
\caption{The histogram of eigenvalues and the smoothed densities of eigenvalues
         for two smoothing parameters for a block-matrix $C_{i,j}$ defined in 
         in item (\ref{it:CorrelMat2}), sec. \ref{sec:simul}.
\label{fig:EigenValDen}.} 
\end{figure}

Now we compute the estimator of the correlation matrix
(\ref{eq:estimator}), diagonalize it and compute the smoothed density of eigenvalues
$\rho_\Lambda^E(\lambda)$ from formula (\ref{eq:DensEigSmooth})
with $\epsilon$ of the order of the mean level spacing between consecutive eigenvalues. 
The results are shown in Fig. \ref{fig:EigenValDen}.
Having obtained the preliminary estimates of parameters $p_\xi$ and $\kappa_\xi$
we carry out the rest of the computation, as described at the end of the previous section,
and obtain the ``cleaned'' correlation matrix $c^{clean}$. 
We quantify the ``goodness of cleaning'' by comparing the mean deviations 
$\left| c^{clean} - C \right|$ and $\left| c - C \right|$.

\begin{figure}
\centerline{\psfig{figure=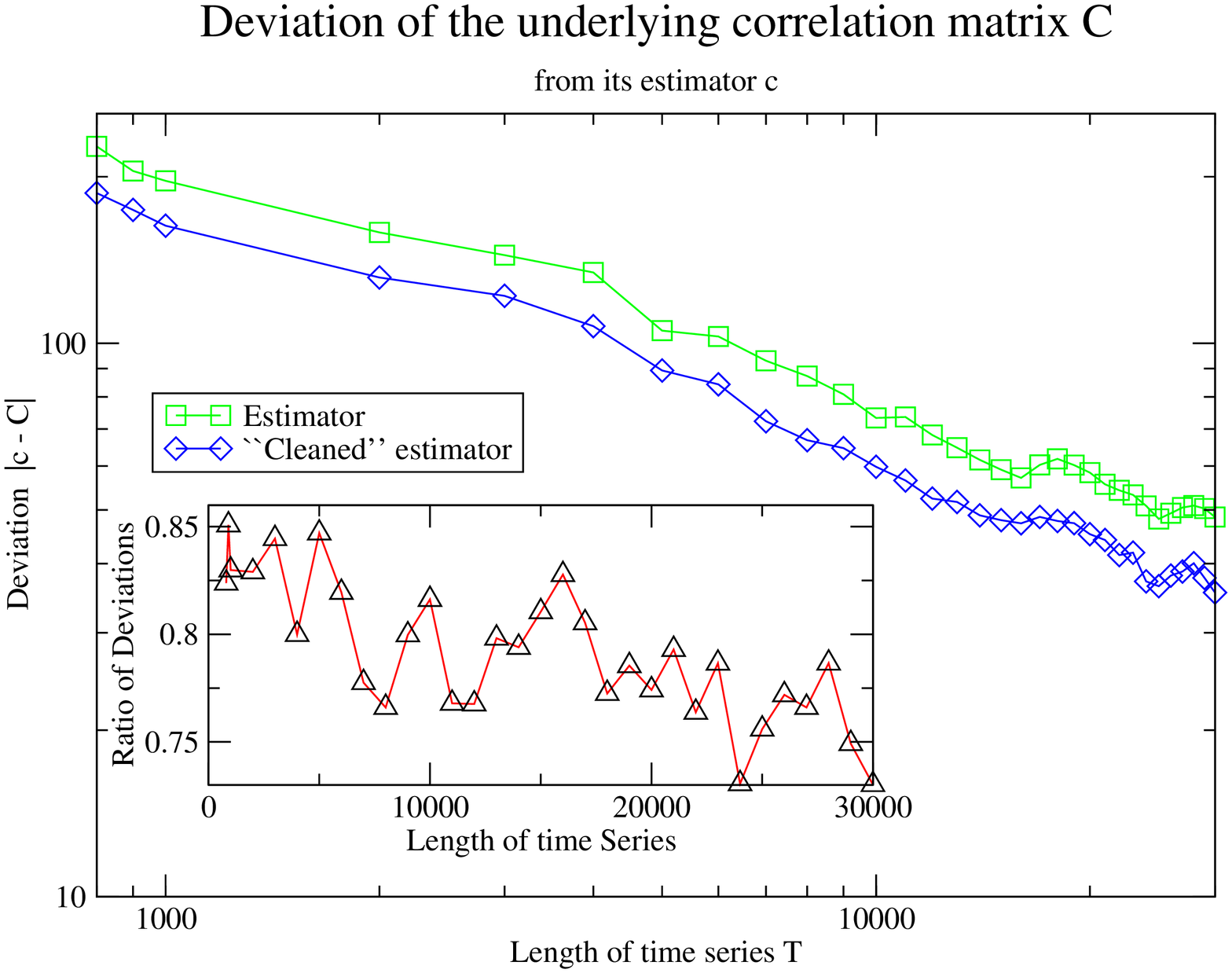,width=0.75\textwidth,angle=-90}}
\caption{Results of a test of the cleaning algorithm for $N = 100$ 
correlated Gaussian stochastic processes for the underlying correlation matrix 
in item (\ref{it:CorrelMat1}), sec. \ref{sec:simul}.
Plotted are deviations $\delta = \sum_{i,j} \left| C_{i,j} - c_{i,j} \right|$ 
between the underlying correlation matrix $C$ 
and the estimator of it $c$($c^E$) before and after applying the ``cleaning procedure''
respectively. The inset includes a ``degree of cleaning'', i.e. ratio of the deviations
after and before applying the denoising procedure.
The ``degree of cleaning'' improves with the length of time series.
\label{fig:TestDenoise}} 
\centerline{\psfig{figure=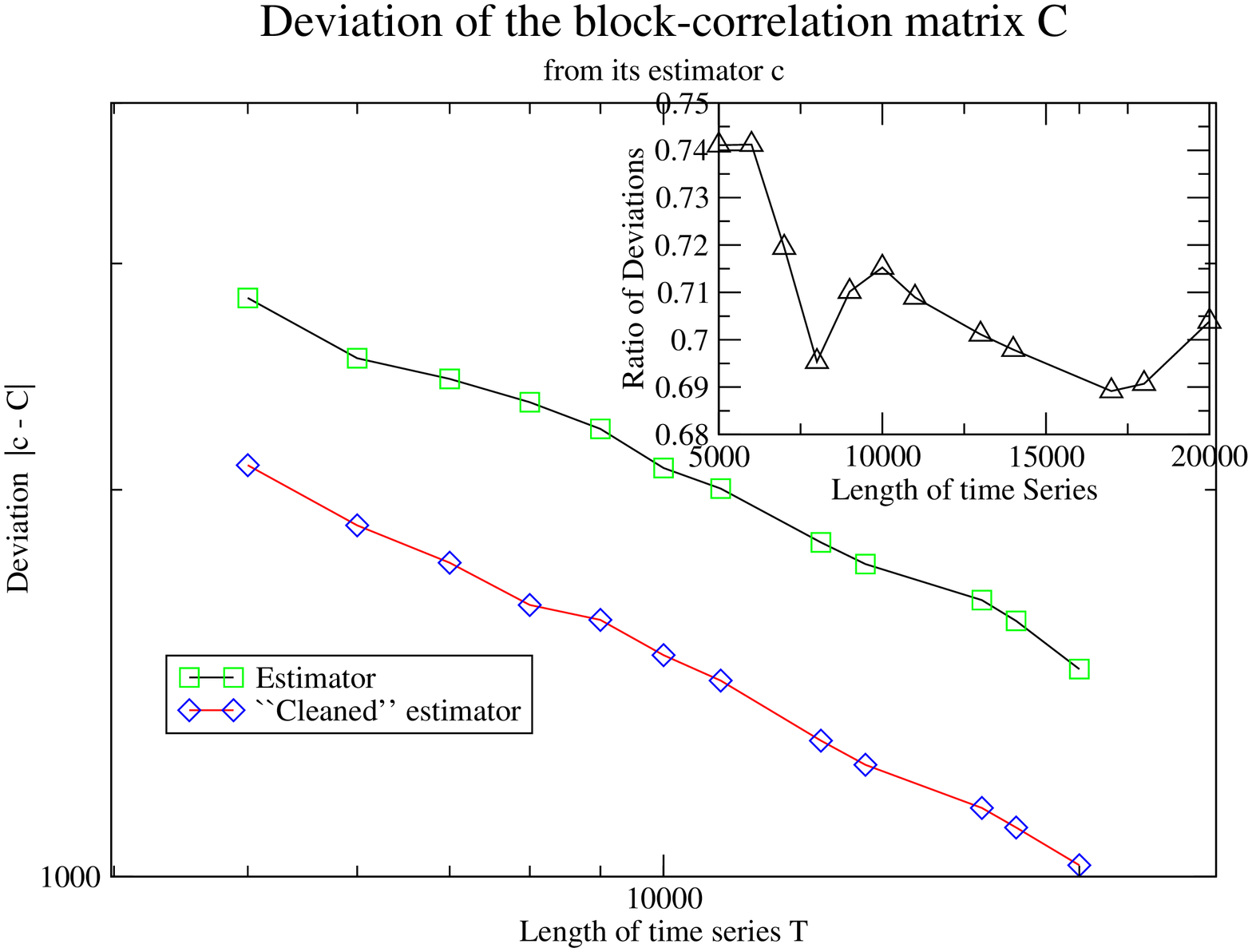,width=0.75\textwidth,angle=-90}}
\caption{The same as in Fig. \ref{fig:TestDenoise} but for the correlation matrix 
in item (\ref{it:CorrelMat2}), sec. \ref{sec:simul}. The dimension is now $N= 493$.
\label{fig:TestDenoise1}} 
\end{figure}

The dependence of the  ``goodness of cleaning'' on the length of time series $T$
is shown in Figs. \ref{fig:TestDenoise},\ref{fig:TestDenoise1}.
We can see two desirable features in the figures.
Firstly the deviation $\left| c - C \right|$
of the estimator from the underlying matrix decreases with increasing 
length of time series. This has nothing to do with the cleaning procedure 
and is only a manifestation of the fact that the estimator $c_{i,j}$
approaches the estimated matrix $C_{i,j}$ monotonously as 
$N/T \longrightarrow 0$.
Secondly the ``goodness of cleaning'' improves with decreasing $N/T$ 
(see inset in Figs. \ref{fig:TestDenoise},\ref{fig:TestDenoise1}).
Indeed it drops from $0.85$($0.74$) to $0.74$($0.7$)
in Figs \ref{fig:TestDenoise} and \ref{fig:TestDenoise1} respectively.
We can, therefore, say that for Gaussian processes the noise filtering
algorithm removes 15\%--26\% and 26\%--30\% noise 
$N/T = 0.1--0.01$ for the both correlation matrices tested in 
in items (\ref{it:CorrelMat1}) and (\ref{it:CorrelMat2}), 
sec. \ref{sec:simul} respectively.

Finally a remark with respect to the choice of the underlying 
correlation matrix $C_{i,j}$.
Since the magnitudes of the eigenvalues of $c_{i,j}$ for financial correlation matrices
differ considerably we need to take the underlying 
correlation matrix with the same property.
We require $C_{i,j}$ to have three distinct eigenvalues 
$\lambda_1 \ll \lambda_2 \ll \lambda_3$,
with degeneracies $N-7$, $4$ and $3$ respectively,
such that $\lambda_2/\lambda_1 = 50$ and $\lambda_3/\lambda_1 = 100$. 
Since the trace of the correlation matrix is $N$ there are certain constraints 
imposed on the $\lambda$'s.
Now, we take two choices of $C_{i,j}$, namely:
\begin{enumerate}
\item $C_{i,j} = f(i-j)$ with $f(i)$ being a Fourier transform of a piece-wise
      constant function with three distinct values. \\
      Here $C = 100^{i-j} +  \sum_{k=1}^{49} 2 {\tilde f}(k) \cos\left\{ 2\pi k (i-j)/N \right\}$
      where 
      \begin{equation}
      {\tilde f}(k) = \left\{ \begin{array}{rr}
                               1 &  \quad\mbox{if}\quad 1  \le k \le 46 \\
                              50 &  \quad\mbox{if}\quad 47 \le k \le 48 \\
                             100 &  \quad\mbox{if}\quad 49 \le k \le 50
                              \end{array}  \right.
      \end{equation}
      and $f(100-k) = f(k)$.
      The matrix $C$ which has three distinct eigenvalues 
      $\left( \lambda_1 = 0.1686340641, 50 \lambda_1, 100 \lambda_1 \right)$  
      with degeneracies $\left( 93, 4, 3 \right)$ respectively\label{it:CorrelMat1}.
\item A block-matrix that consists of ${\mathcal S}$ blocks $\tilde{C}_\xi$ 
      of dimensions $n_\xi$
      for $\xi=1,\ldots,{\mathcal S}$ on the diagonal and zeroes elsewhere.
      The blocks itself have a structure
      \begin{equation}
      \tilde{C}_\xi = \left( \begin{array}{ccccc}
                               1 & \mathcal{C} & \mathcal{C} & \ldots & \mathcal{C} \\
                               \mathcal{C} & 1 & \mathcal{C} & \ldots & \mathcal{C} \\
                               \mathcal{C} & \mathcal{C} & 1 & \ldots & \mathcal{C} \\
                               \vdots                 \\
                               \mathcal{C} & \mathcal{C} & \mathcal{C} & \ldots & 1
                              \end{array}  \right)  
      \end{equation} where $-1 \le \mathcal{C} \le 1$.
      This choice is dictated by the fact there may exist groups of stocks on the market,
      which have the property that two stocks are only correlated if they belong to the same 
      group. This means that 
      $\left< X_{\xi,i} X_{\eta,j} \right> = \delta_{\xi,\eta} \mathcal{C}_{i,j}$ 
      for $i,j=1,\ldots,n_\xi$ and $\xi,\eta=1,\ldots,S$. 
      In the following we choose $\mathcal{C}_{i,j} = \mathcal{C}$  for simplicity
      and also because it renders the smallest eigenvalue of all block matrices
      to be the same.
      The minimal dimension of such matrix $C_{i,j}$ that has
      three distinct eigenvalues $\lambda_1,\lambda_2,\lambda_3$
      with  degeneracies $N-7$, $4$ and $3$ and for which
      $\lambda_2/\lambda_1 = 50$ and $\lambda_3/\lambda_1 = 100$
      is $N = 493$.
      The matrix contains four blocks of dimension $n_1= 49$
      and three blocks of of dimension $n_5 = 99$ and the parameter $\mathcal{C}$ 
      in each block is $\mathcal{C}= 1/2$\label{it:CorrelMat2}.
\end{enumerate}

\section{Results for financial data \label{sec:results}}
Having tested the algorithm on Gaussian stochastic processes we now apply
it to daily financial data from the Standard and Poor's 500 stock group.
Here we check if the eigenvalues $\lambda$ and their degeneracies $\mathcal{D}$, 
which we obtain
as a result of the minimization procedure described above,
correspond to eigenvalues obtained by \citet{Laloux} and \citet{Bouchaud}.
What one did there was to order the eigenvalues in ascending order, divide them into
groups (there were usually only two groups, the ``noisy part'' containing most eigenvalues
and one separated eigenvalue much larger than the rest) and replace
the eigenvalues from the group by one degenerated eigenvalues equal to the mean over 
eigenvalues from the group (in order to conserve the trace of the correlation matrix).
In other words the group of eigenvalues 
$\lambda_{i_1}, \lambda_{i_2}, \ldots, \lambda_{i_p}$ was replaced by one eigenvalue
$\left< \lambda \right> = \left( \sum_j \lambda_{i_j} \right)/p$ 
with a degeneracy $p$. 

\begin{figure}
\centerline{\psfig{figure=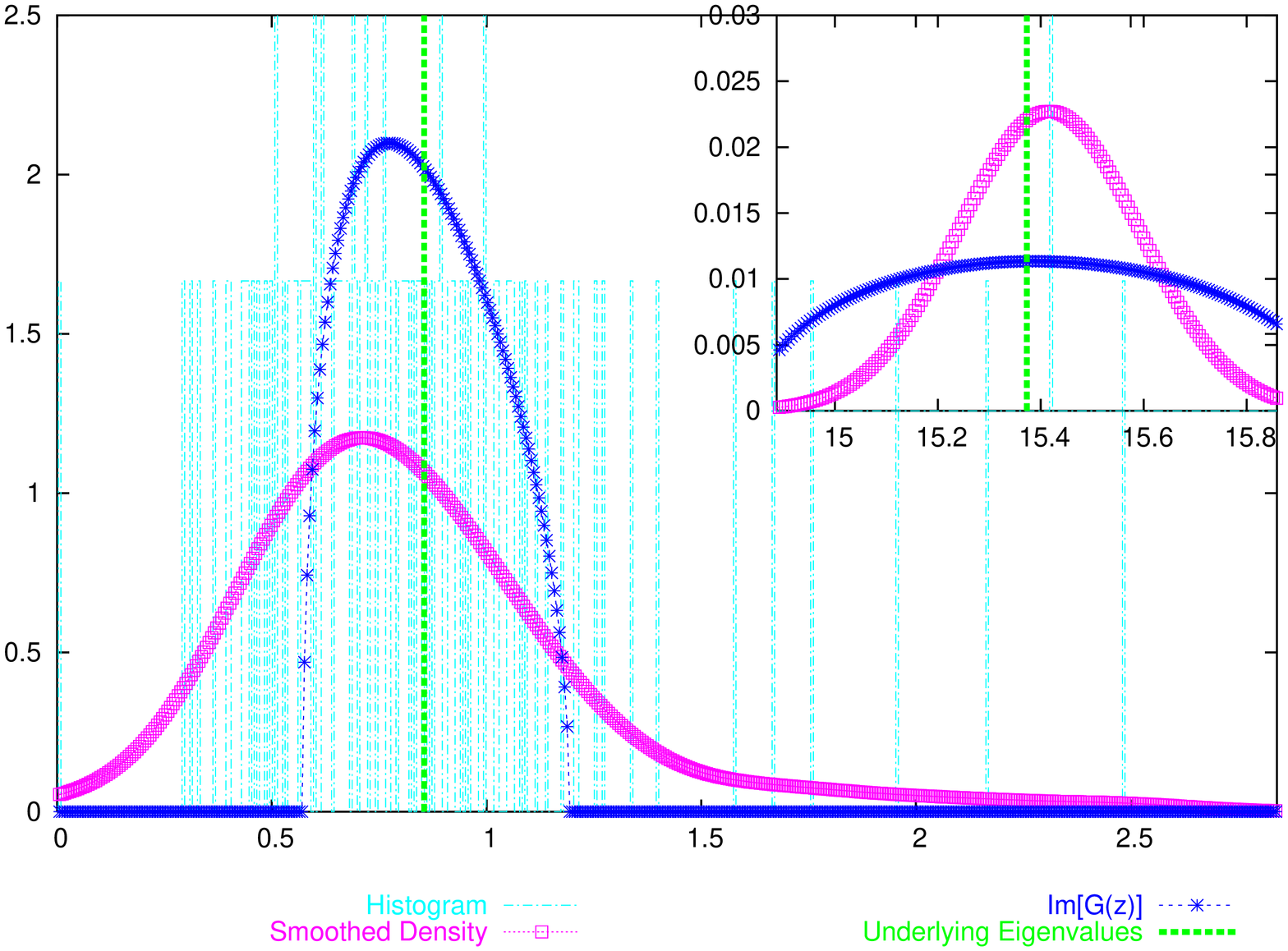,width=0.75\textwidth,angle=-90}}
\caption{The histogram of the density of eigenvalues of the correlation 
         matrix of the first 
         hundred stocks 
         AA--CMS
         from the Standard and Poor's 500 group, 
         the smoothed density $\rho_\epsilon(\lambda)$, 
         the imaginary part $\rho_\Lambda(\lambda)$ 
         of the Greens function 
         and the eigenvalues of the ``cleaned'' 
         correlation matrix $C_{i,j}$.
         The empirical moments are
         $m^{E} = \left(1.000,1.000,3.314,38.064,567.738,8713.976,134235.952,2069070.046,\right)$,
         the moments corresponding to the resolvent are
         $M = \left(1.000,1.000,3.120,37.267,564.756,8701.178,134235.871,2071710.416\right)$.
         \label{fig:Correl100}}
\centerline{\psfig{figure=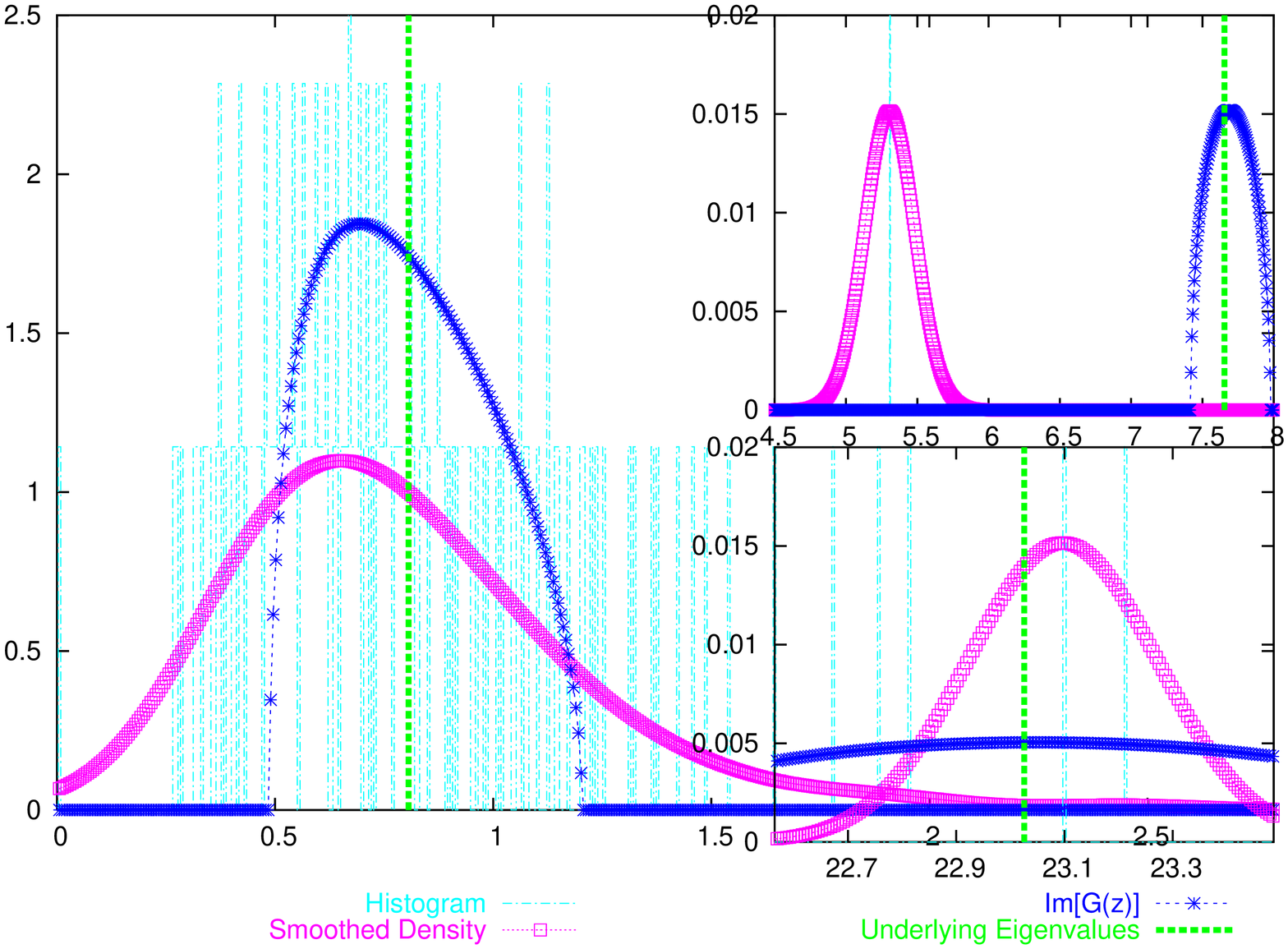,width=0.75\textwidth,angle=-90}}
\caption{The same as in figure Fig. \ref{fig:Correl100} but for the first 150 S\&P500 stocks.
         \label{fig:Correl150}}
\end{figure}

\begin{figure}
\centerline{\psfig{figure=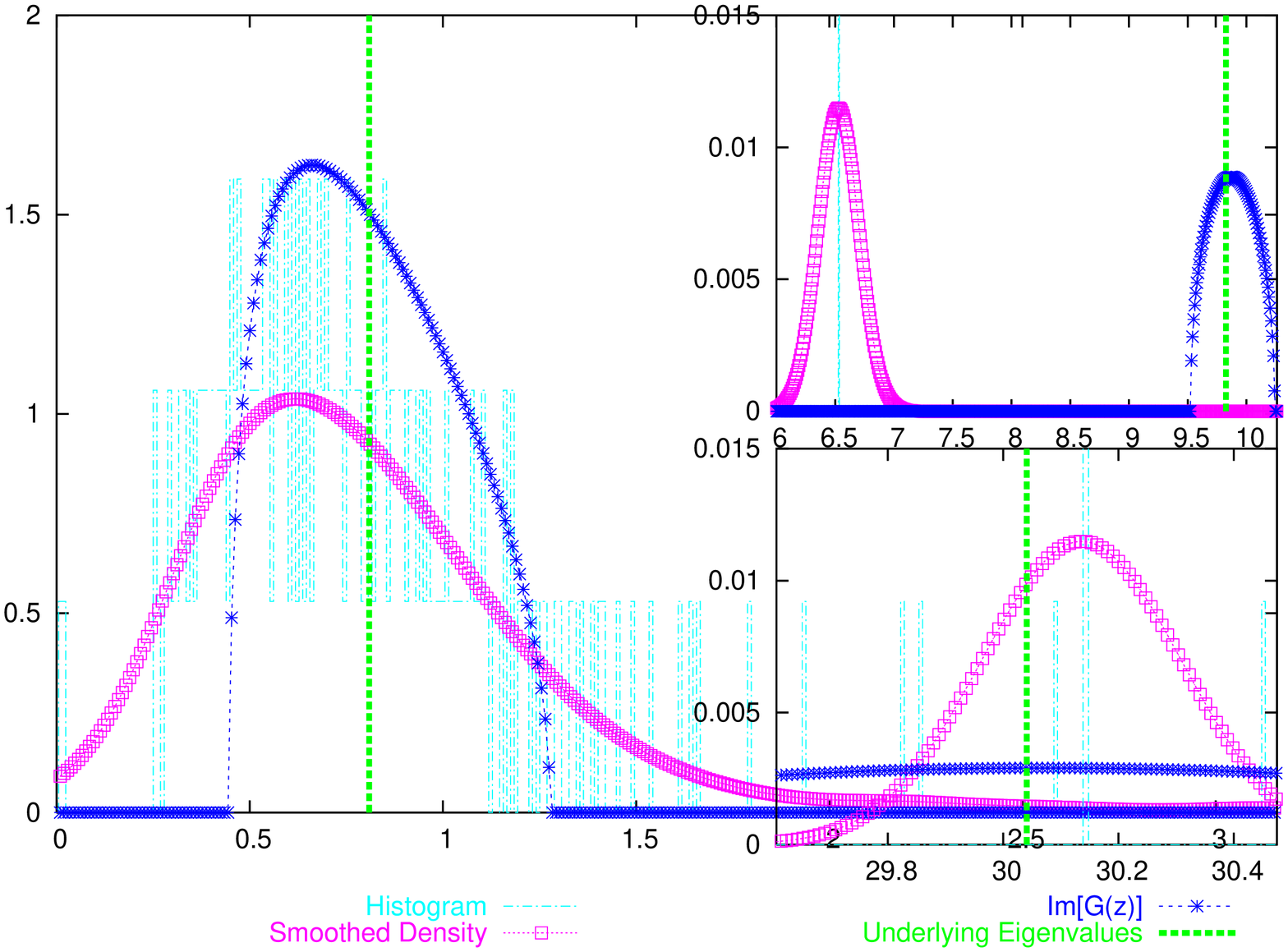,width=0.75\textwidth,angle=-90}}
\caption{The same as in figure Figs. \ref{fig:Correl100},\ref{fig:Correl150} 
         but for the first 200 S\&P500 stocks.
         \label{fig:Correl200}}
\centerline{\psfig{figure=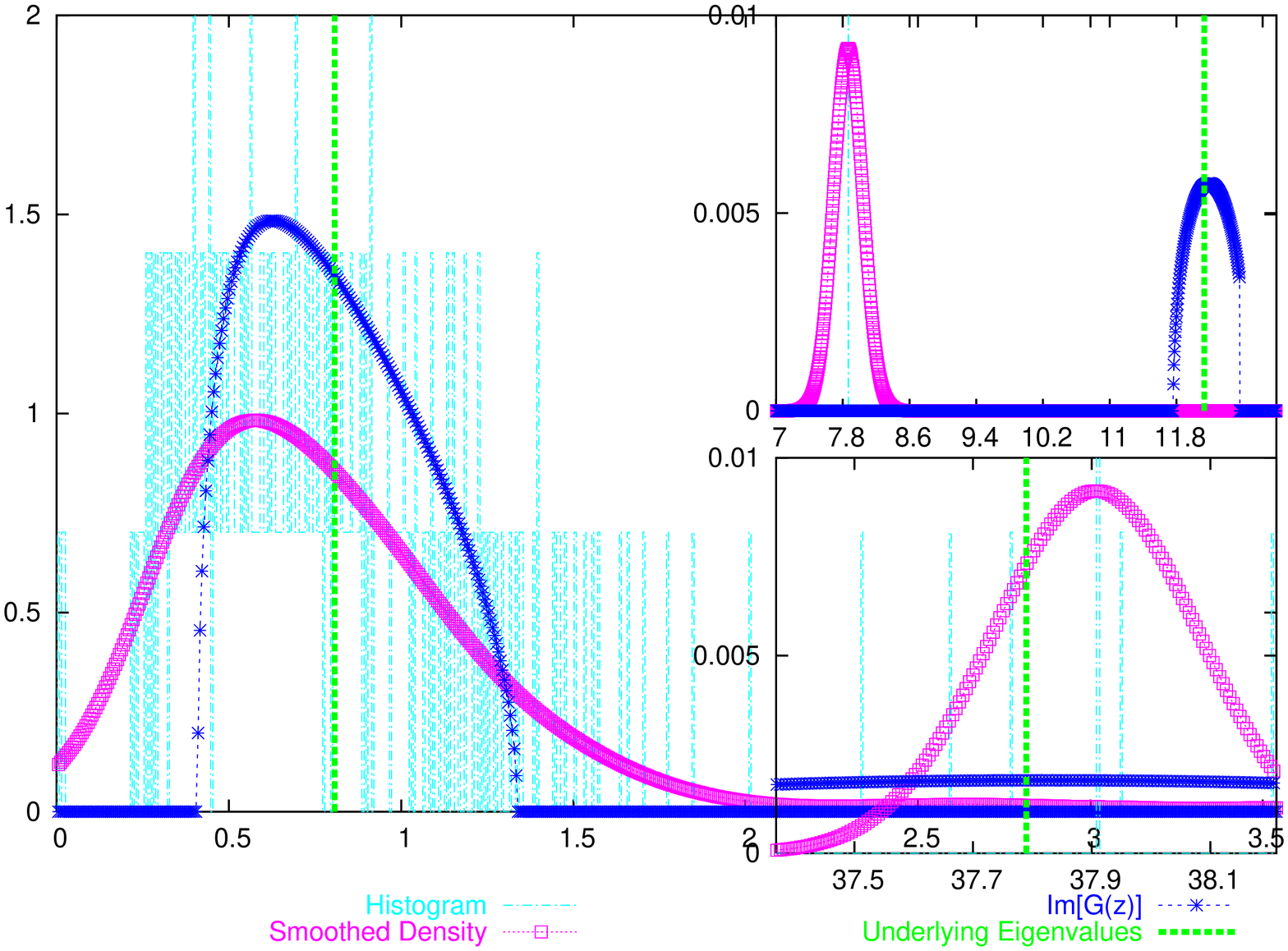,width=0.75\textwidth,angle=-90}}
\caption{The same as in figure Figs. \ref{fig:Correl100},\ref{fig:Correl150}, 
         \ref{fig:Correl200} 
         but for the first 250 S\&P500 stocks.
         \label{fig:Correl250}}
\end{figure}

\begin{figure}
\centerline{\psfig{figure=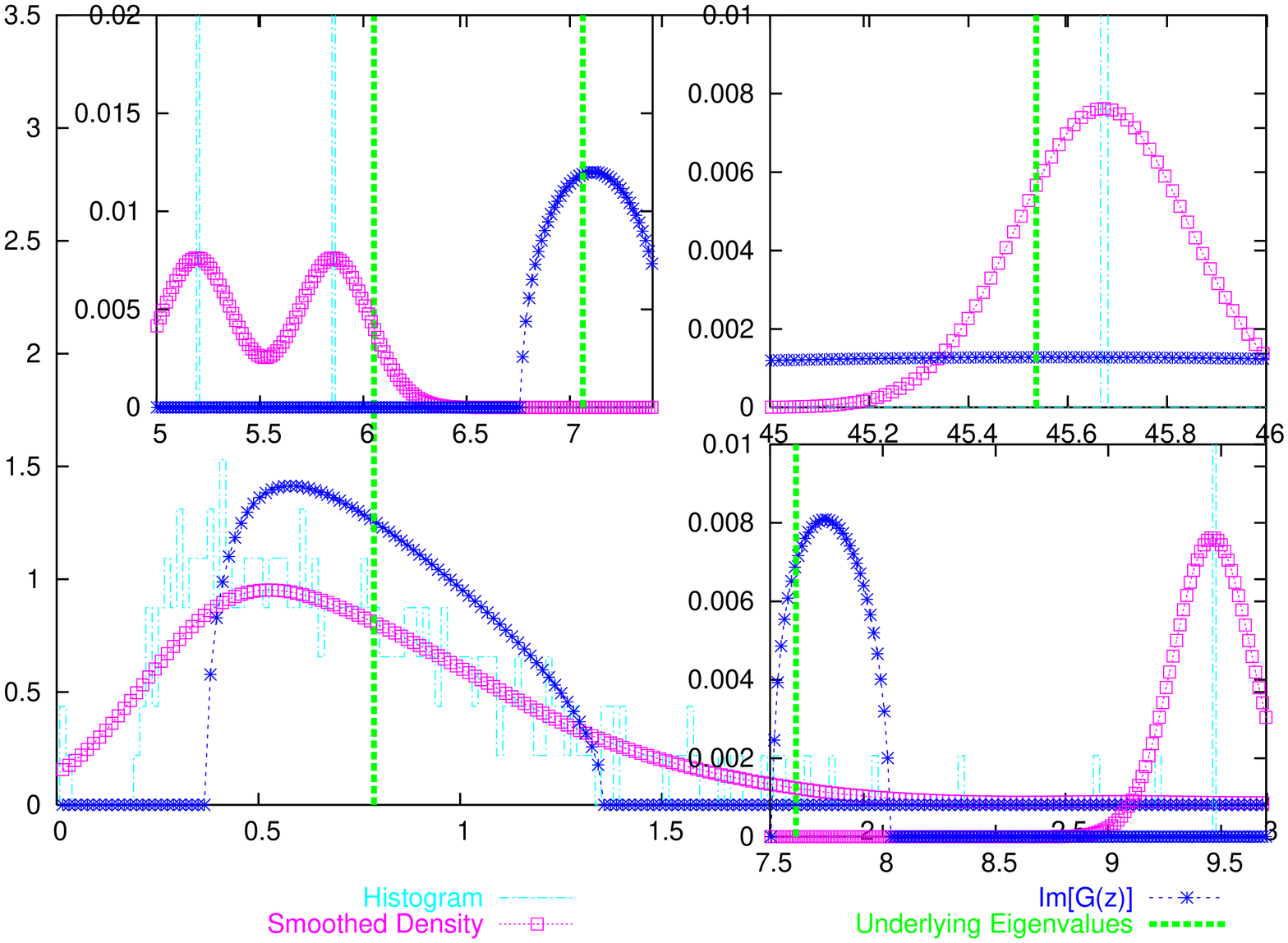,width=0.75\textwidth,angle=-90}}
\caption{The same as in figure Figs. \ref{fig:Correl100}--\ref{fig:Correl250}
         but for the first 300 S\&P500 stocks.
         \label{fig:Correl300}}
\centerline{\psfig{figure=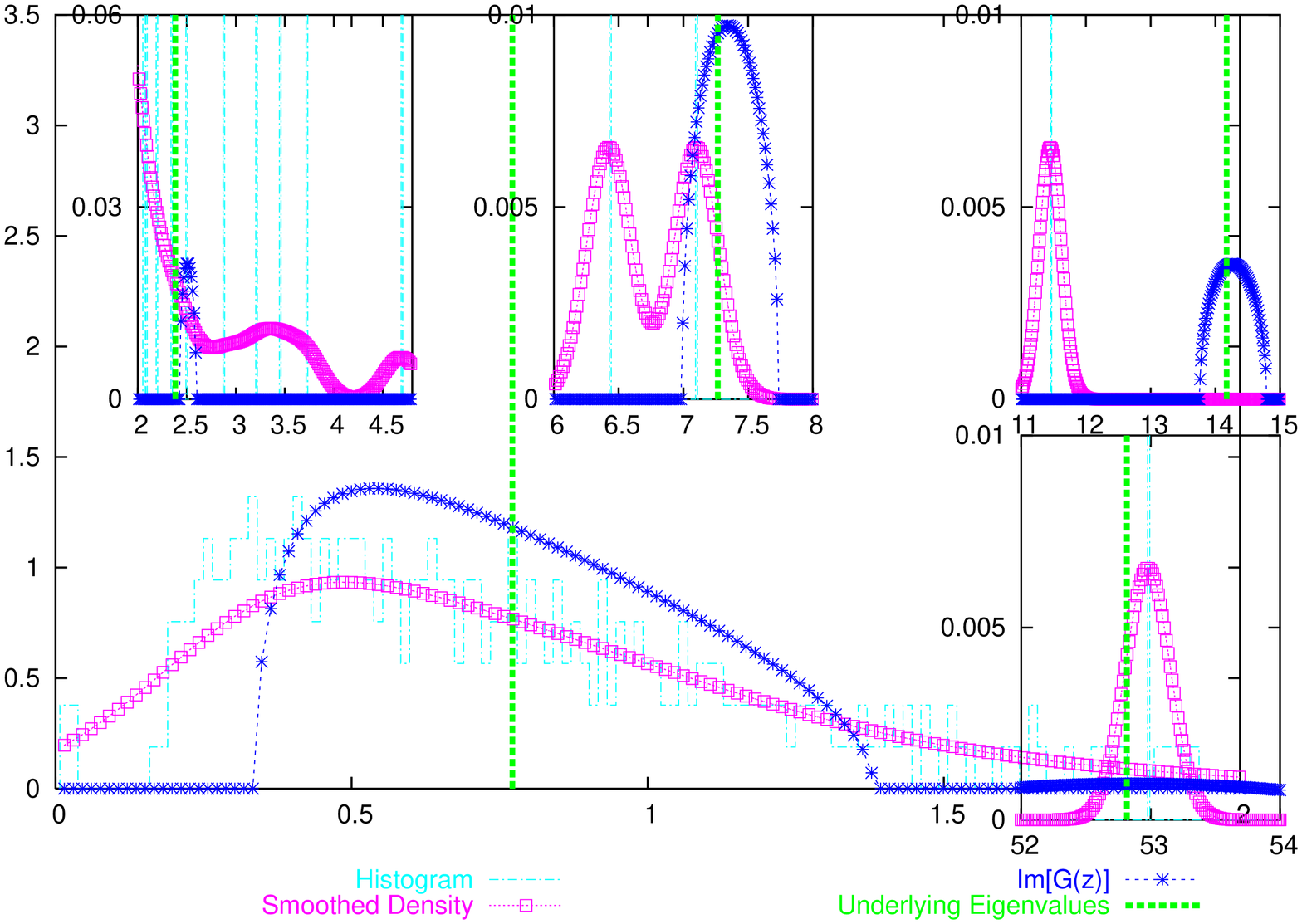,width=0.75\textwidth,angle=-90}}
\caption{The same as in figure Figs. \ref{fig:Correl100}--\ref{fig:Correl300}
         but for the first 350 S\&P500 stocks.
         \label{fig:Correl350}}
\end{figure}

\begin{figure}
\centerline{\psfig{figure=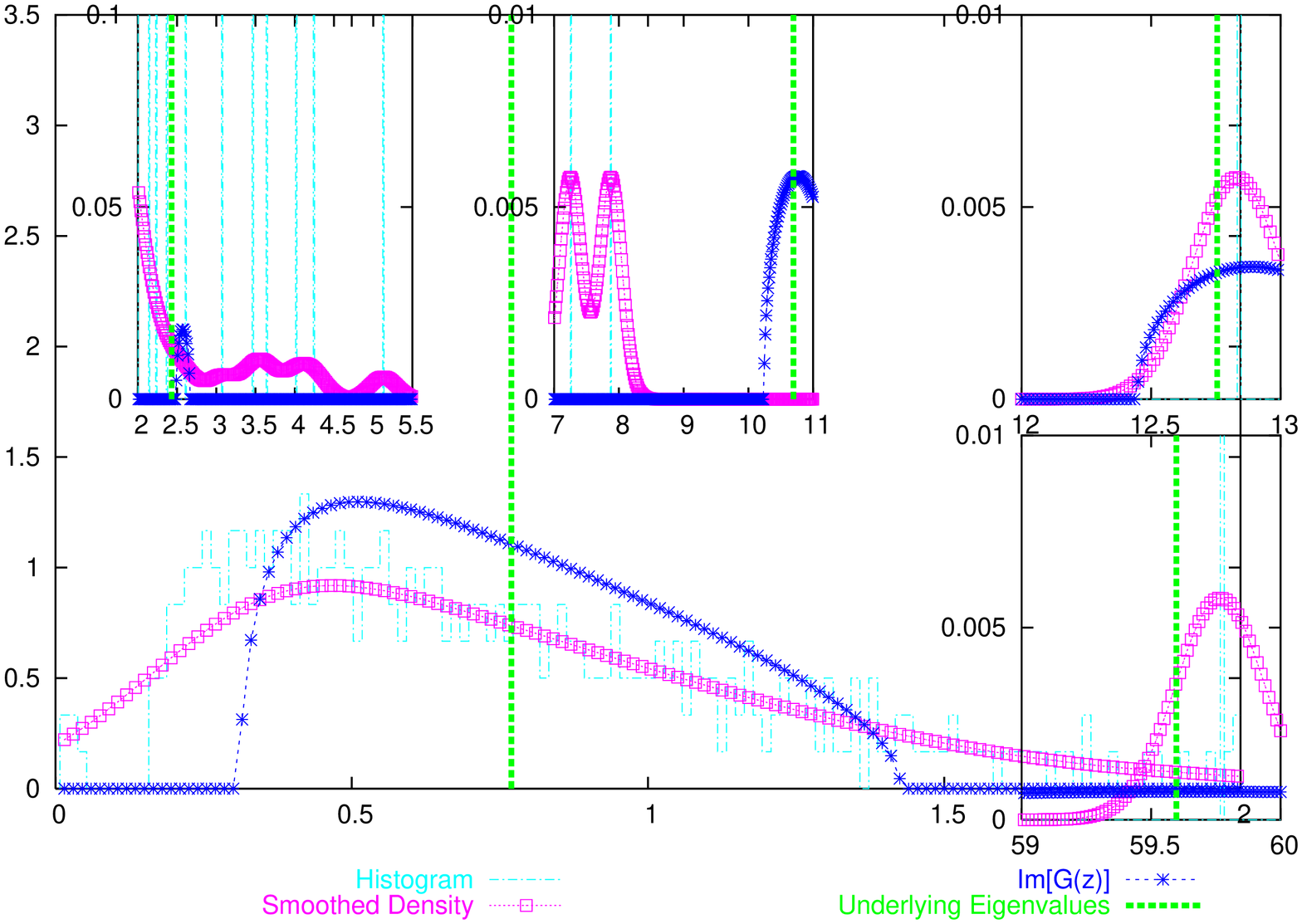,width=0.75\textwidth,angle=-90}}
\caption{The same as in figure Figs. \ref{fig:Correl100}--\ref{fig:Correl350}
         but for the first 400 S\&P500 stocks.
         \label{fig:Correl400}}
\centerline{\psfig{figure=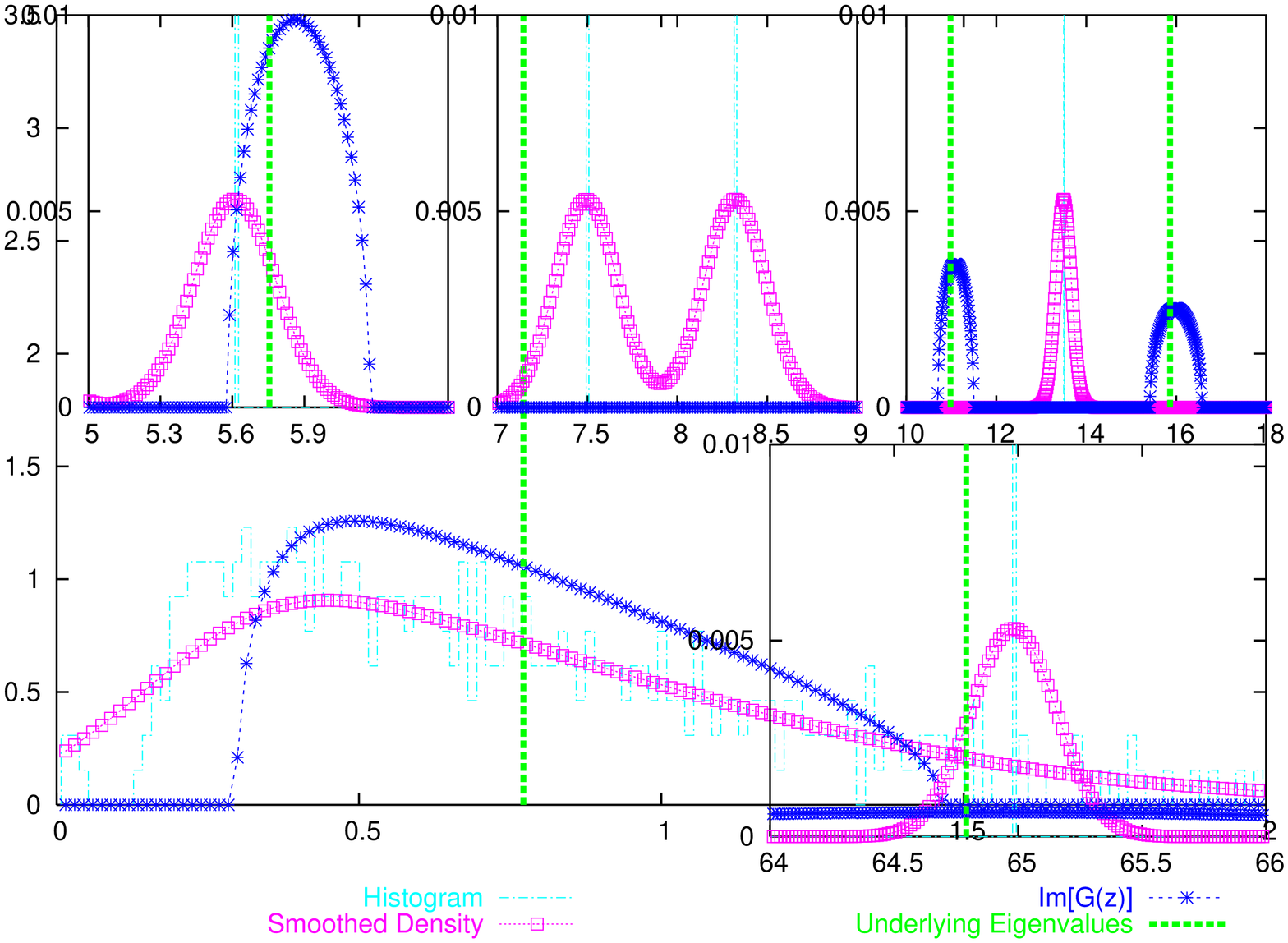,width=0.75\textwidth,angle=-90}}
\caption{The same as in figure Figs. \ref{fig:Correl100}--\ref{fig:Correl400}
         but for the first 435 S\&P500 stocks.
         In contrast to the previous cases here the underlying eigenvalues
         do not match the midpoints of the bands but even though the deviation
         $\delta = |\rho_\Lambda(\lambda) - \rho_\Lambda^E(\lambda)|$ is minimal 
         (compare Table \ref{tab:EigenValLoc}).
         \label{fig:Correl435}}
\end{figure}

We have performed computations for groups of stocks with a fixed length $T = 3000$ days
and with a growing number of stocks $N$.
The optimal eigenvalues and their degeneracies are shown in Table \ref{tab:EigenValLoc}.
As we can see the algorithm is stable, i.e. it produces results which are close to that 
$\left< \lambda \right>$ obtained by \citet{Laloux} for different values of $N$. 
In other words we have shown that the minimization of deviations between
the spectrum of eigenvalues of the measured estimator of cross-correlations and 
the spectrum of eigenvalues of a certain random matrix 
amounts, approximately, to the procedure by \citet{Laloux} and \citet{Bouchaud}, 
i.e. to replacing each group of eigenvalues by one single, degenerated eigenvalue
whose value is equal to the mean over the group.
However, as we can see in Table \ref{tab:EigenValLoc}, 
the eigenvalues $\lambda$ and $\left< \lambda \right>$ do not match exactly.
This is due to the fact that sometimes it is difficult to define a band
(see Figs. \ref{fig:Correl100}--\ref{fig:Correl435})
in the spectrum of the measured estimator. This depends on the smoothing parameter
$\epsilon$ in (\ref{eq:DensEigSmooth}).
Moreover, it is not quite clear that the minimization routine will lead us to the 
global minimum. If there are more than $\mathcal{P} = 2$ bands then it is 
quite likely for the algorithm to get stuck in a local minimum, which will 
be hard to distinguish from the global minimum.
Indeed, in the last row in Table \ref{tab:EigenValLoc} the underlying eigenvalues $\lambda$'s
and the mean eigenvalues $\left< \lambda \right>$ do not match but even though
the deviation $\delta$ is of the same order than in previous rows.

From Table \ref{tab:EigenValLoc} we see that the minimization was convergent 
in a reasonable time (the number $\mathcal{N}$ of function evaluations was reasonable) 
even if there were several parameters to be optimized. 

\begin{table}
\begin{tabular}{llllllll}
$N$ &  \makebox[0.12\textwidth]{
       \begin{minipage}{0.12\textwidth}
       Range of stocks 
       \end{minipage}} 
    & $\lambda$ ($\mathcal{D}$) 
    & $\left< \lambda \right>$ ($\mathcal{D}$) 
    & $\left| c - c^{clean} \right|$
    & $\delta$ 
    & $\mathcal{P}-1$
    & $\mathcal{N}$ \\ \hline
$100$                 & \makebox[0.12\textwidth]{
                        \begin{minipage}{0.12\textwidth}
                        AA--CMS
                        \end{minipage}} 
                      & \makebox[0.14\textwidth]{
                        \begin{minipage}{0.14\textwidth}
                        0.8548(99) \\
                        15.3729(1)   
                        \end{minipage}}            
                      & \makebox[0.14\textwidth]{
                        \begin{minipage}{0.14\textwidth}
                        0.8544(99) \\
                        15.4158(1)   
                        \end{minipage}}    
                      & 0.02895
                      & 0.0902   & 1 & 49 \\ \hline
$150$                 & \makebox[0.12\textwidth]{
                        \begin{minipage}{0.12\textwidth}
                        AA--EP
                        \end{minipage}} 
                      & \makebox[0.14\textwidth]{
                        \begin{minipage}{0.14\textwidth}
                        0.8062(148) \\
                        7.6541(1)   \\
                        23.0255(1) 
                        \end{minipage}}            
                      & \makebox[0.14\textwidth]{
                        \begin{minipage}{0.14\textwidth}
                        0.8216(148) \\
                        5.3063(1)   \\
                        23.0968(1) 
                        \end{minipage}}   
                      & 0.02732
                      & 0.0236 & 2 & 134 \\ \hline
$200$                 & \makebox[0.12\textwidth]{
                        \begin{minipage}{0.12\textwidth}
                        AA--HLT
                        \end{minipage}} 
                      & \makebox[0.14\textwidth]{
                        \begin{minipage}{0.14\textwidth}
                        0.8088(198) \\
                        9.8260(1)   \\
                        30.0408(1) 
                        \end{minipage}}            
                      & \makebox[0.14\textwidth]{
                        \begin{minipage}{0.14\textwidth}
                        0.8249(198) \\
                        6.5286(1)   \\
                        30.1357(1) 
                        \end{minipage}}   
                      & 0.02666
                      & 0.0228 & 2 & 100 \\ \hline
$250$                 & \makebox[0.12\textwidth]{
                        \begin{minipage}{0.12\textwidth}
                        AA--LTD
                        \end{minipage}} 
                      & \makebox[0.14\textwidth]{
                        \begin{minipage}{0.14\textwidth}
                        0.8068(248) \\
                        12.1353(1)   \\
                        37.7897(1) 
                        \end{minipage}}            
                      & \makebox[0.14\textwidth]{
                        \begin{minipage}{0.14\textwidth}
                        0.8235(248) \\
                        7.8638(1)   \\
                        37.9071(1) 
                        \end{minipage}}   
                      & 0.02661
                      & 0.0248 & 2 & 154 \\ \hline
$300$                 & \makebox[0.12\textwidth]{
                        \begin{minipage}{0.12\textwidth}
                        AA--NYT
                        \end{minipage}} 
                      & \makebox[0.14\textwidth]{
                        \begin{minipage}{0.14\textwidth}
                        0.7862(296) \\
                        7.0621(2) \\
                        7.6146(1) \\
                        45.5366(1) 
                        \end{minipage}}            
                      & \makebox[0.14\textwidth]{
                        \begin{minipage}{0.14\textwidth}
                         0.7899(296) \\
                         5.5244(2)\\
                         9.4625(1)\\
                         45.6709(1)
                        \end{minipage}}   
                      & 0.02049 
                      & 0.0295 & 3 & 124 \\ \hline
$350$                 & \makebox[0.12\textwidth]{
                        \begin{minipage}{0.12\textwidth}
                        AA--SCH
                        \end{minipage}} 
                      & \makebox[0.14\textwidth]{
                        \begin{minipage}{0.14\textwidth}
                         0.7713(345) \\
                         2.3814(1)\\
                         7.2660(2)\\
                         14.1737(1)\\
                         52.8160(1)
                        \end{minipage}}            
                      & \makebox[0.14\textwidth]{
                        \begin{minipage}{0.14\textwidth}
                         0.7750(345) \\
                         4.6842(1)\\
                         6.7622(2)\\
                         11.4487(1)\\
                         52.9780(1)
                        \end{minipage}}   
                      & 0.01904
                      & 0.0112 & 4 & 293 \\ \hline
$400$                 & \makebox[0.12\textwidth]{
                        \begin{minipage}{0.12\textwidth}
                        AA--UCL
                        \end{minipage}} 
                      & \makebox[0.14\textwidth]{
                        \begin{minipage}{0.14\textwidth}
                        0.7692(395) \\
                        2.4292(1)\\
                        10.6966(2)\\ 
                        12.7553(1)\\ 
                        59.5976(1)\\
                        \end{minipage}}            
                      & \makebox[0.14\textwidth]{
                        \begin{minipage}{0.14\textwidth}
                        0.7776(395) \\
                        5.1295(1) \\
                        7.5616(2) \\
                        12.8304(1) \\
                        59.7666(1)
                        \end{minipage}}   
                      & 0.01969
                      & 0.0099 & 4 & 259 \\ \hline
$435$                 & \makebox[0.12\textwidth]{
                        \begin{minipage}{0.12\textwidth}
                        AA--ZION
                        \end{minipage}} 
                      & \makebox[0.14\textwidth]{
                        \begin{minipage}{0.14\textwidth}
                        0.7718(430) \\
                        15.8633(1) \\
                        5.7546(2) \\
                        10.9826(1) \\
                        64.7905(1) \\
                        \end{minipage}}            
                      & \makebox[0.14\textwidth]{
                        \begin{minipage}{0.14\textwidth}
                        0.7793(430) \\
                        5.6098(1)\\
                        7.9069(2)\\
                        13.4956(1)\\
                        64.9812(1)\\
                        \end{minipage}}   
                      & 0.02611
                      & 0.0101387 & 4 & 409 \\ \hline

\end{tabular}
\caption{Eigenvalues and degeneracies $\lambda$ and $\mathcal{D}$
         of the underlying correlation matrix 
         mean eigenvalues 
        $\left< \lambda \right>$ in the consecutive bands,
        the sum of relative differences of moments $\delta$ of the empirical
        correlation matrix and the corresponding ``stochastic'' matrix, 
        i.e. that of the Wishart matrix ensemble,
        the deviation $\left| c - c^{clean} \right|$ 
        of the ``cleaned'' estimator from the measured estimator,
        the number of parameters $\mathcal{P} - 1$ optimized
        and the number of function evaluations $\mathcal{N}$
        in the downhill-simplex method needed to achieve a tolerance of $10^{-6}$.
        Data correspond to daily historical times series ($T = 3000$ days) 
        of different groups of stocks of the
        Standard and Poor's 500 group of shares. \label{tab:EigenValLoc}}
\end{table}

\section{Optimal portfolio\label{sec:OptPortf}}
\begin{figure}
\centerline{\psfig{figure=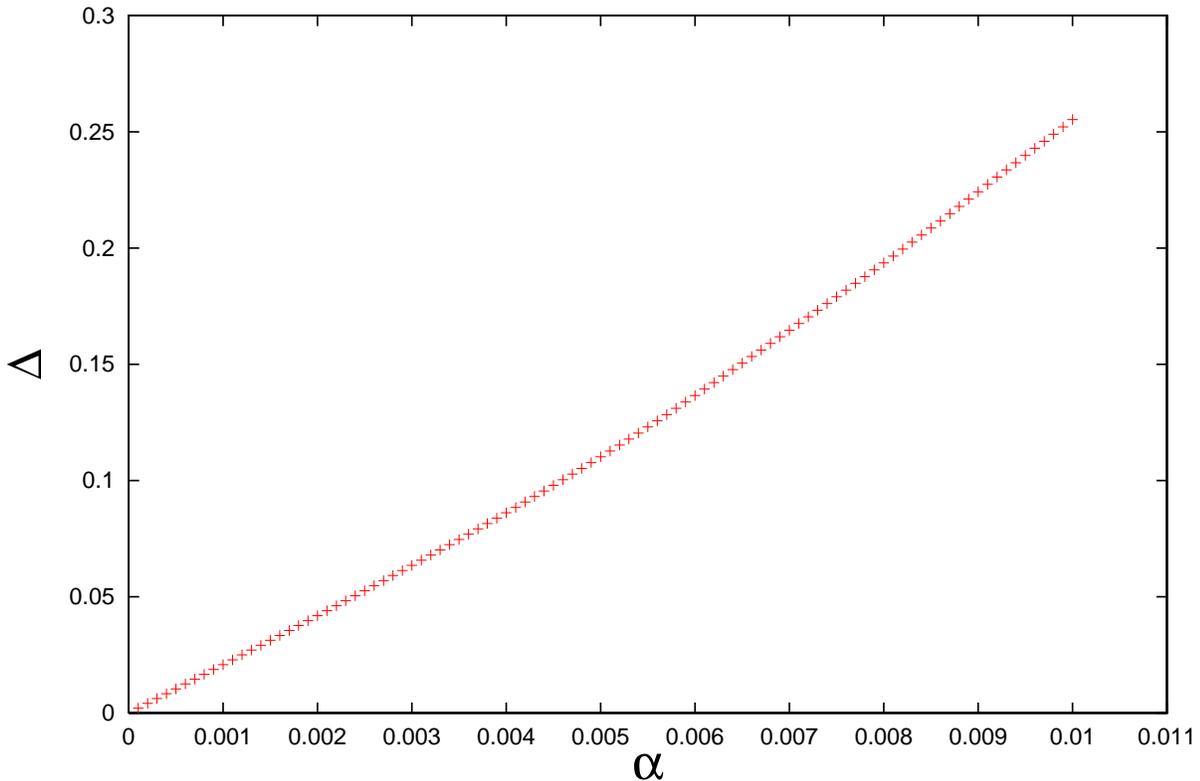,width=0.8\textwidth,angle=-90}}
\caption{The average relative variation in the weights of stocks (Markovitz-weights)
under addition of independent Gaussian noise with mean zero and variance $\alpha^2$
to the correlation matrix defined in item (\ref{it:CorrelMat2}), 
sec. \ref{sec:simul}.\label{fig:MarkowitzWeightsVariat}}
\end{figure}

According to \citet{Markovitz} the optimal portfolio $P=\{p_1,\ldots,p_N\}$ 
of independent, correlated Gaussian stochastic processes,
that ensures
the smallest possible risk $D_P = \sum_{i,j} p_i p_j C_{i,j}$
subject to the constraint $\sum_i p_i = 1$ reads:
\begin{equation}
p_i = \frac{\sum_j C^{-1}_{i,j}}{\sum_{i,j} C^{-1}_{i,j}}
\label{eq:PortfolioWeigths}
\end{equation}
Practical applications of this recipe are limited because the weights
$p_i$ are very sensitive to perturbations of the matrix elements $C_{i,j}$.
We have tested the sensitivity in the following simple experiment.
Having perturbed the correlation matrix $C_{i,j}$
with Gaussian noise with mean zero and variance $\alpha^2$,
i.e 
\begin{equation}
C^{\mathcal{P}}_{i,j} = C_{i,j} + \alpha \mathcal{R}_{i,j}
\end{equation}
where $\mathcal{R}_{i,j} = Normal(0,1)$ we measure the
average relative variation 
\begin{equation}
\Delta = \frac{2}{N} \sum_{i=1}^{N} 
\frac{\left| p_i^{\mathcal{P}} - p_i\right|}
     {\left( \left|p_i\right| + \left|p_i^{\mathcal{P}}\right|\right)}
\end{equation}
in the Markovitz' weights
defined in (\ref{eq:PortfolioWeigths}).
Here $p_i$ and $p_i^{\mathcal{P}}$ are related to the matrices $C_{i,j}$
and $C^{\mathcal{P}}_{i,j}$ respectively.  
The results are shown in Fig. \ref{fig:MarkowitzWeightsVariat}.
As we can see an addition of only 1\% noise ($\alpha$ = 0.01) makes 
the Markovitz' weights change already by 25\% ($\Delta$ = 0.25).
On the other hand in Fig. \ref{fig:TestDenoise1} we see that
the estimator of correlations $c_{i,j}$ 
differs from the underlying correlation matrix $C_{i,j}$ by approximately
0.8\%--1.2\% for $N/T$ = 0.05--0.1.
In sec. \ref{sec:simul} we concluded that the algorithm
removes on average 25\% noise once the ration $N/T \le 0.1$.  
Therefore the noise contamination of the ``cleaned'' correlation matrix
reduces to 0.6\%--0.9\% and this gives us, 
from Fig.\ref{fig:MarkowitzWeightsVariat}, a decrease of 
the variation $\Delta$ in Markovitz weights 0.13--0.17
and 0.22--0.3 in case of the smallest and largest values of $N/T$ respectively.
We see that the relative percentage variation in the Markovitz-weights
after applying the ``cleaning'' procedure is about $(0.17-0.13)/0.17$ $\simeq$ 25\%. 

%

\newpage

\section{Conclusions}
We have proved that the qualitative cross-correlations' ``cleaning'' procedure 
suggested by \cite{Bouchaud} and \citet{Laloux} corresponds to a 
numerical minimization problem,
namely to the problem of minimizing deviations between the
eigenvalue spectrum of the measured cross-correlations 
and the spectrum of a random matrix. 
Moreover, we have shown qualitatively that if the ratio $N/T \le 0.1$
the algorithm filters out roughly 25\% noise from Gaussian 
distributed stochastic processes, which corresponds to a 
percentage correction of the Markovitz' weights by roughly 25\%.
Quantification of this statement will be desirable.

Future work will be devoted to developing
an analysis of cross-correlations between stocks in a non-Gaussian market
where distributions of returns exhibit power-law tails.     

%



\end{document}